\documentclass[twocolumn,pre,showpacs,showkeys]{revtex4}
\usepackage{graphicx,color,amssymb,amsfonts,amsmath,float}
\newcommand{\figforce}{%
\begin{figure}[htbp]
\includegraphics[width=\linewidth,clip]{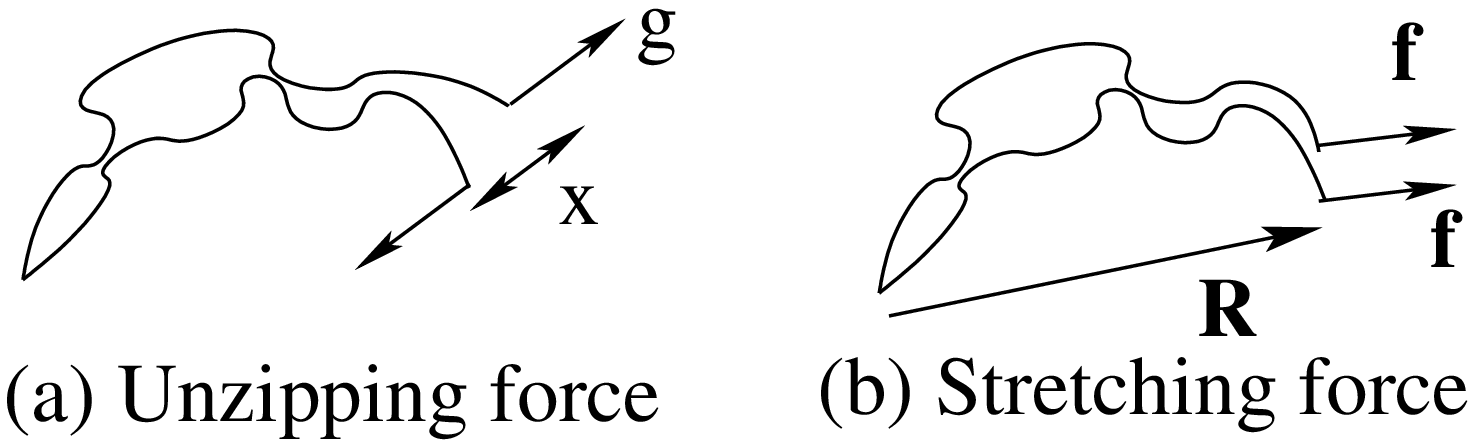}
\caption{ Two   types of  external forces on a DNA. (a)
    Unzipping force where the ends of the two strands are pulled in
    opposite directions. (b) Stretching force where the two ends are
    pulled in the same direction. }
\label{fig:1}
\end{figure}
}%

\newcommand{\figkappa}{%
\begin{figure}[b]
\includegraphics[width=\linewidth,clip]{kappa.eps}
\caption{The unzipping phase boundary near the melting point for
  different values of $\kappa$, the shape exponent.
}
\label{fig:3}
\end{figure}
}%

\newcommand{\figextr}{%
\begin{figure}[htbp]
\includegraphics[width=0.5\linewidth,clip]{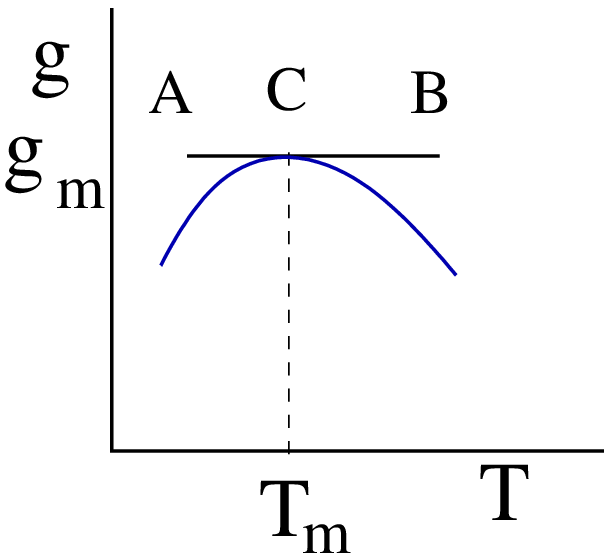}
\caption{A maximum at C=$(T_m,g_m)$ in the $T$-$g$  phase boundary. 
A path ACB 
   sees  certain special features at C.  A zipped phase at
$T=T_m$ can be unzipped by a force $g_m$ without any discontinuity in entropy.
}
\label{fig:4}
\end{figure}
}%

\newcommand{\figxg}{%
\begin{figure}[htbp]
\includegraphics[width=\linewidth,clip]{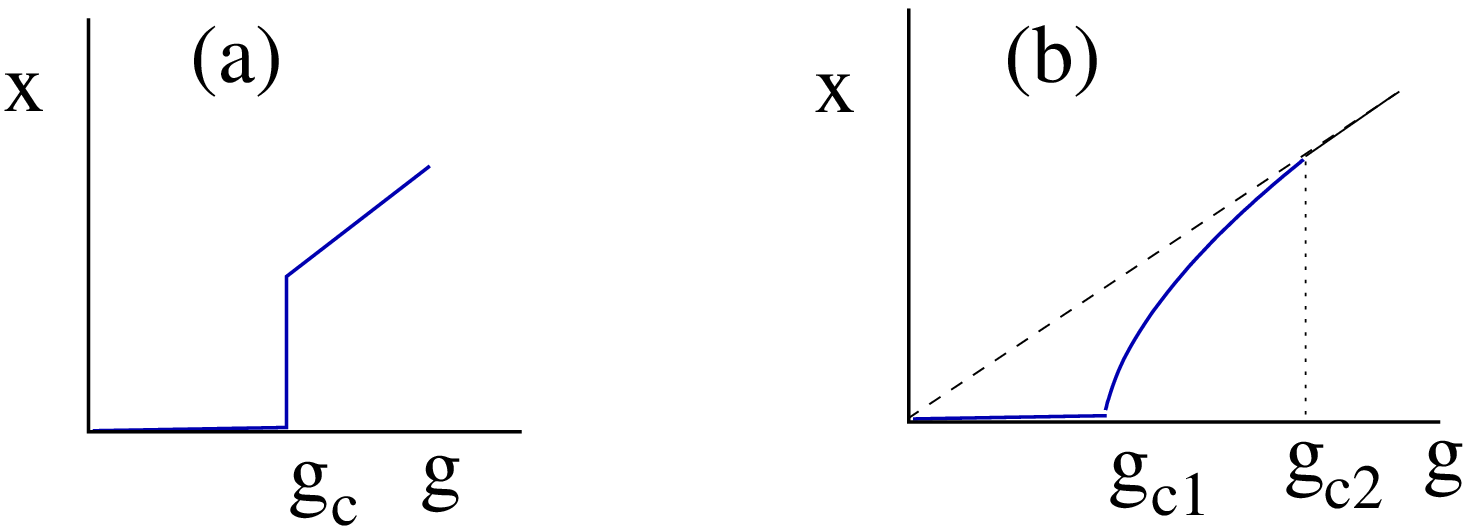}
\caption{Possible isotherms (constant $T$). (a) A first-order
  unzipping transition at $g=g_c$. There is a jump in $x$. (b) Two
  continuous transitions at $g=g_{c1}$ and $g=g_{c2}$.  The force does
  not affect the DNA for $g<g_{c1}$ but penetrates and modifies the
  bound state continuously from $g>g_{c1}$ to $g<g_{c2}$.  The unbound
  or stretched denatured phase occurs for $g>g_{c2}$.}
\label{fig:5}
\end{figure}
}%

\newcommand{\figsp}{%
\begin{figure}
\includegraphics[width=\linewidth,clip]{CS_vert.eps}
\includegraphics[width=\linewidth,clip]{CS_hori.eps}
\caption{Specific heat for the $d=1$ exactly solvable model with no
  crossing.  The phase diagram is similar to Fig. \ref{fig:4} with
  $T_m=0.904642475...$ and $g_m= 1.358806498...$ in the units chosen.
  $C(T_m,g)$ [in (a)] and $S(T_m,g)$ [in (b)] vs $g$.  In (a) there is
  a discontinuity as we go through the peak of the phase boundary
  vertically by changing $g$ at $T=T_m$. In (b), we see that the
  entropy is continuous at the transition point. There is no latent
  heat.  In (c) $C(T,g_m)$ vs $T$ and in (d) $S(T,g_m)$ vs $T$ are
  shown for a fixed $g=g_m$, i.e., along the horizontal line of Fig.
  \ref{fig:4}.  
  The zipped phase occurs only at one point without any signature
  elsewhere.  This contribution is shown by a star.  
The entropy remains continuous and analytic throughout
  but the specific heat has one extra discontinuous point at $T=T_m$.
}
\label{fig:6}
\end{figure}
}%

\begin{document}

\title{Thermodynamic relations for DNA phase transitions}
\author{Poulomi Sadhukhan}
\email{sadhukhan@theorie.physik.uni-goettingen.de}
\affiliation{Institut f\"ur Theoretische Physik, Universit\"at G\"ottingen, Friedrich-Hund-Platz 1, 37077 G\"ottingen, Germany
}
\author{Somendra M. Bhattacharjee}
\email{somen@iopb.res.in }

\affiliation{Institute of Physics, Bhubaneswar 751 005, India}

\begin{abstract}
  The force induced unzipping transition of a double stranded DNA is
  considered from a purely thermodynamic point of view.  This analysis
  provides us with a set of relations that can be used to test
  microscopic theories and experiments.  The thermodynamic approach is
  based on the hypothesis of impenetrability of the force in the
  zipped state.  The melting and the unzipping transitions are
  considered in the same framework and compared with the existing
  statistical model results.  The analysis is then extended to a
  possible continuous unzipping transition.
\end{abstract}

\maketitle

\section{Introduction}
\label{sec:introduction}
To read the genetic information encoded in the base sequence, hidden
in the helical structure of a DNA, it is necessary to break the
hydrogen bonds of the base pairs\cite{watson}.  The mechanism for
doing so is the unzipping by a force \cite{smb1,smb2} of a double
stranded DNA (dsDNA), or a thermal melting \cite{melting}.  In the
melting transition, the hydrogen bonds of base pairing are broken by
thermal energy, while in the unzipping transition, it is by a pulling
force at one end of the DNA.  In both cases, the strands remain
intact.

While there is a long history of experimental studies of the melting
transition\cite{melting}, the investigations of the unzipping
transition or responses to external forces are of more recent
origin\cite{danilo2004}.  Pioneering calorimetric studies were done
over a large range of temperature ($T$) from 2K to 400K under
different solution conditions\cite{mrevlishvili}.  So far as force is
concerned, isotherms of DNA, like the response under a force have been
obtained in many different types of single molecule
experiments\cite{kumarli,smith1996,ritort2010}.  Nevertheless, calorimetry
in presence of a force is still not available.

It is known from various theoretical models that, for both melting and
unzipping, the nature of the transition depends on the aspects of the
DNA captured in a
model\cite{kumarli,smb1,marensmb,MTM2001,orland2001,kafri2002,kapri2004,girikumar,joyeux,senooverstret,giri2007,kapri2009,zhang1997}.
Any natural DNA, because of its large length, is expected to show the
characteristic features of the transitions;  however, the
situation is not so clear on the experimental front.  In-vitro
experiments are generally restricted to short chains.  Consequently,
very little is known about the role of sequence variation, e.g., as
seen across species, vis-a-vis the true transition behaviour expected of
long chains.  As a matter of fact, we even lack a clear experimental
answer about the order of the melting transition \cite{munoz2010,peyrard2011}.

The co-operativity in melting comes from the entropy ($S$) of the DNA
through the correlations introduced by the strands as long
polymers\cite{degennes,smbflory}.  The unzipping transition is due to
the competition between the pairing of the strands and the stretching
of the unbound strands\cite{smb1,marensmb,MTM2001,orland2001}.  The
work done in stretching the free polymers provides the cost of
unpairing the strands.  This cost at zero temperature is only the
pairing energy, but, because of entropy, the critical unzipping force
vanishes as one approaches the melting temperature.  The thermodynamic
conjugate pair for the transition is $g$, the unzipping force, and
$x$, the separation of the two strands at the point of application of
force (see Fig. \ref{fig:1}).  
 It transpires that gross quantities
like the entropy, the specific heat, and the response function for
force, are the relevant thermodynamic quantities to study, especially
as the transition point is approached.  The advantage in the
thermodynamic approach is that the results obtained are valid under
quite general conditions without getting into the microscopic details
of DNA.

%%%%%%
\figforce
%%%%%%

Besides the conjugate pairs ($T, S$), and $(g,x)$, there could be
other types of external forces, e.g.,   isotropic hydrostatic or osmotic
pressure affecting the volume of the polymer, and  a stretching
force (${\bf f}$) that distorts and elongates the chain.  Although
there are evidences of hydrostatic or osmotic pressure affecting
protein-DNA
interaction\cite{amiri,wiltonBDNA,chalikian,takahashi2013,erijman,robinson94},
there is only a very weak effect on the melting of DNA.  In contrast,
a stretching force may lead to an ``overstretching'' transition where
the length of the DNA increases by a factor of 1.7
\cite{smith1996,overstretch,senooverstret,maiti}.  
Whether it is an
equilibrium (meaning thermodynamic) transition is still debated.

The unzipping
transition was first established in a continuum model in
Ref. \cite{smb1,smb2}. It  was also proved by studying the
dynamics of pulling in Ref. \cite{sebastian} and by several exactly
solvable models\cite{marensmb,MTM2001,kapri2004}.
Various aspects of the unzipping transition, and in that context the
corresponding melting transition, have been studied.   These include
the effects of randomness  in
interaction, or force \cite{nelson,allahv,tamm,random},
semiflexibility\cite{kierfeld}, and finite length \cite{kapri2006,szhao}.
Many details of the transition have also been studied, like various
distributions\cite{shikha}, temperature dependence\cite{navin},
different types of noise\cite{kapri2005,puiman},   role of
ensembles\cite{alexander}.  The dependence of melting on the nature of the space
has also been studied via the choice  of different fractal
lattices\cite{kumar1993,jm2012,jm2014,izv2007,izivic2008}, 
showing  the possible  variations in the melting transition. The
mapping of the DNA melting  problem to a quantum problem revealed the connection
between the bubble entropy of DNA and the quantum transition
\cite{ps2012}, 
and to the Efimov-physics\cite{pal2013}.  
Biological applications have also 
been considered, especially the motion of the interface or the
Y-fork\cite{smb2010,li2006}.

Our purpose in this paper is to consider the melting and the unzipping
transitions from a purely thermodynamic point of view, without any
consideration of any microscopic models.  This way we derive the
relevant thermodynamic relations applicable to these transitions.
Obviously such predictions are independent of the microscopic details.
{\it{Our basic hypothesis is that the bound  double-stranded state
of DNA    is a thermodynamic phase that does not allow penetration of an
    unzipping force.}}  In other words, the linear response function
for a weak force in the zipped state is strictly zero.  (The words
``zipped''  and  ``bound'' are to be used as synonymous.)
We start with the definitions and standard relations in terms of the
DNA variables in Sec. \ref{sec:therm-descr}.  The case of a first
order unzipping transition is discussed in Sec.
\ref{sec:first-order-unzipp}.  Here we consider the case of no
penetration of force in the bound state.  In other words the bound
state remains the same till the critical unzipping force is reached.
The thermodynamic predictions are then compared, in
Sec. \ref{sec:exact-results}, with the known exact solutions in
certain class of models.  Although all theoretical studies based on
simple coarse-grained models predict a first-order unzipping
transition, there is a proposal that local penetration of forces may
lead to a continuous transition\cite{mixed}.  Thermodynamics does not
rule out any continuous unzipping transition, but, instead,  allows a 
different phase with partial penetration of force.  A thermodynamic
analysis of such a case of a continuous transition is discussed in
Sec.  \ref{sec:cont-unzipp-trans}.  In this case we assume that for a
range of force $g_{c1}\le g \le g_{c2}$, there is a change in the DNA
bound state by the external force.  A few details can be found in the
Appendices. The additions of other forces like hydrostatic pressure
and a stretching force are discussed in Appendix A.  The relevant
Maxwell relations for DNA unzipping are listed in Appendix B.  The
specific heat relation for a continuous transition can be found in
Appendix C.

\section{Thermodynamic description}
\label{sec:therm-descr}
Our main concern is in the unzipping transition and therefore we
restrict ourselves to the $g$ and $x$ pair.  In absence of any other
information, we may allow both the unzipping and the melting
transition to be either first order or continuous.  Both cases are
discussed here.
 
What makes the problem different from others is the fact that the
unzipping force does not affect the bound state for small forces.  In
fact only other system that shows similar thermodynamic relations is a
superconductor with the Meissner phase not allowing the external
magnetic field to penetrate\cite{supercon}.  In that analogy, a
parallel scenario for DNA would be the case where the force penetrates
for an intermediate range of force, leading to a continuous
transition\cite{mixed}.

One may consider two mutually exclusive situations, either $g$ or $x$
is fixed. These correspond to the two possible ensembles in
the statistical mechanical approach. The fixed-force case described by
the Helmholtz free energy $F(T,x)$ and the
fixed-distance ensemble, described by the Gibbs free energy $G(T,g)$.
These are in addition to the usual  canonical (fixed-$T$)
and micro-canonical (fixed-$S$) ensembles. 
The free energies are given by 
\begin{subequations}
\begin{eqnarray}
  \label{eq:1}
  F(T,x)&=&U-T\;S,\\
   G(T,g)&=&U-T\;S\;-\;g\;x =F-g\;x,
\end{eqnarray}
\end{subequations}  
where $U$ is the internal energy.  Henceforth, we use $F,G,U,S$ to
mean the corresponding quantities per monomer or base pair.  The
differential form for $G$ is
\begin{equation}
  dG=-S\;dT -x\; dg.\label{eq:3}  
\end{equation}
It is possible to extend the thermodynamic formulation to include
other external forces.  Some details may be found in Appendix A.

By integrating Eq. \eqref{eq:3} at constant temperature, one gets the Gibbs
free energy at a force $g$ as
\begin{equation}
  \label{eq:8}
  G(T,g)=G(T,0)-\int_0^g x\; dg.
\end{equation}
This form is valid for equilibrium with $x=x(g)$ as the equilibrium
isotherm of a DNA and is used extensively in this paper.  The formula
for work done in Eq. \eqref{eq:8} is different from the mechanical
definition of work ($\int g dx$).  A justification is as follows.  In
a nonequilibrium situation, to change the force from zero to $g$, the
work done on the DNA is $\int_0^g x dg$ for a trajectory.  For
example, an instantaneous change in force would require a work $w=xg$
if the distance remains fixed at $x$.  Then the histogram
transformation in statistical mechanics gives us the free energy
difference as \cite{ps2010}
\begin{equation}
  \label{eq:39}
  G(T,g)-G(T,0)=-k_BT \ln \;\langle \exp(-\beta w)\rangle,
\end{equation}
where $\beta=(k_BT)^{-1}$, and the angular bracket indicates averaging
over all possible trajectories starting with the equilibrium
distribution at zero force.  In this particular case of instantaneous
increase, the averaging is over all values of $x$ with the equilibrium
probability distribution $P_{T,g}(x)$ at the initial force.  For an
infinitesimal increment $dg$, from $g$ to $g+dg$, we may expand
$\exp(-\beta x dg)\approx 1-\beta (x \;dg)$.  Therefore, for small
$dg$, the equivalent of Eq. \eqref{eq:39} is
\begin{eqnarray}
  \label{eq:40}
    \Delta G(T,g)&=&-k_BT \ln \left(1 -  \beta\int dx \; (x\; dg)\;
      P_{T,g}(x)\right )\nonumber\\
      &=& x(g)\ dg,
\end{eqnarray}
where the average value of $x$ is denoted by $x(g)$.
On successive integration, one recovers the thermodynamic formula of
Eq. \eqref{eq:8} (with no angular bracket).
Incidentally, the mechanical work done in stretching or unzipping has
been used in other contexts too, as, e.g.,  to obtain and use  the
hysteresis around the transition for thermodynamic free energies\cite{kapri2011}
and associated dynamic transitions\cite{gmps,gm2013,qiyi2011}.

The notations we are using are as follows.  The zero force thermal
melting temperature is denoted by $T_c$.  The unzipping transition by
a force $g$ at temperature $T$ takes place at a temperature dependent
force $g=g_c(T)$ so that $g_c(T_c)=0$.

%%%%%%%%%%%%
\figxg
%%%%%%%%%%%%

\section{ First order Unzipping transition}\label{sec:first-order-unzipp}
For the unzipping transition, $G(T,g)$ is continuous across the phase
boundary.  This implies 
\begin{equation}
  \label{eq:41}
G_{\rm z}(T,g_c)=G_{\rm u}(T,g_c),  
\end{equation}
where subscripts  z and u  indicate  the zipped and the unzipped phases.
Eq. \eqref{eq:8} therefore allows us to write
\begin{equation}
  \label{eq:9}
G_{\rm z}(T,0) - G_{\rm u}(T,0)=\int_0^{g_c}  (x_{\rm z}-x_{\rm u})\; dg.
\end{equation}
Here $G_{\rm u}(T,0)$  is
the free energy of the unzipped phase in zero force if it had
existed.  One way of obtaining $G_{\rm u}(T,0)$ is by extrapolation of
the high force free-energy, assuming that the extrapolation is
thermodynamically admissible, or from the free energy of a single
stranded DNA.

It is known that for the first order unzipping transition (Fig. \ref{fig:5}a),
the force does not penetrate the bound state for \mbox{$g<g_c(T)$}. 
As mentioned in the introduction, we
take this as the starting hypothesis in the thermodynamic analysis.
  Therefore, effectively, $x_{\rm z}=0$,
and
\begin{equation}
  \label{eq:7}
  G_{\rm z}(T,g)=G_{\rm z}(T,0), \quad(g\le g_c).
\end{equation}
This equation is valid at $g=g_c$
because of coexistence of phases.  At this point the force-dependent
unzipped phase has the same $G$ as the zipped phase.
In the linear response regime, $x_{\rm u}=\chi_T\; g$ where
$\chi_T$, the extensibility, may be taken  to be a constant. 
(See Appendix B for definitions.)  
Eq. \eqref{eq:9} then simplifies to
\begin{equation}
  \label{eq:20}
  G_{\rm z}(T,0) = G_{\rm u}(T,0) -\frac{1}{2} \chi_T \; g_c^2,
\end{equation}
where the last term is the work   $W(g_c)$.
A more useful  form is obtained by combining  Eqs. \eqref{eq:8}, ~\eqref{eq:7},
and ~\eqref{eq:20}, as
\begin{equation}
  \label{eq:10}
  G_{\rm z}(T,g) = G_{\rm u}(T,g) +\frac{1}{2} \chi_T \; (g^2
  -g_c^2),
\end{equation}
in principle, valid for all $g$.  This  shows that for
$g<g_c$, the zipped phase is more stable than the unzipped one and
vice versa.

\subsection{Entropy}
\label{sec:entropy}
The entropy difference, from Eq. \eqref{eq:20}, (see Appendix B) comes out to be
\begin{eqnarray}
  \label{eq:11}
  S_{\rm z}(T,g_c)-S_{\rm u}(T,g_c)&=& \chi \;g_c(T)\ \frac{\partial
    g_c(T)}{\partial T}\\
                     &=& x(g_c)\ \frac{\partial
    g_c(T)}{\partial T},
\end{eqnarray}
where the second form, a more general one, follows  by noting that
$\partial W(g)/\partial g= x$.  For notational simplicity, we omit the subscripts of 
$\chi$.
The entropy difference is related to the latent heat $L=T (S_{\rm
  z}-S_{\rm u})$ at the transition.  Except for $g=0$, energy is
required to unzip a DNA.  In real situations this energy is supplied
by nonthermal sources like ATP etc.

The continuity of the Gibbs free energy at the unzipping transition point in a
fixed force ($dG_{\rm z}=dG_{\rm u}$ along the phase boundary), gives
the Clausius-Clapeyron equation as 
\begin{equation}
  \label{eq:13}
  \frac{\partial g_c}{\partial T}=\frac{S_{\rm u}-S_{\rm z}}{x_{\rm
      u}-x_{\rm z}},
\end{equation}
where all the quantities on the right hand side are on the phase
boundary.  The impenetrability condition, $x_{\rm z}=0$ with 
the linear response relation $x_{\rm u}= \chi g_c$, yields the entropy
relation of Eq. \eqref{eq:11}.

The sign of the right hand side in Eq. \eqref{eq:11}, i.e., the slope
of the phase boundary, is not 
fixed {\it a priori}.    This is important for identification of the state which
is more ordered.   In a temperature driven transition, the entropy 
increases as one crosses a phase transition line from the low to the
high temperature side.   If the zipped phase is more ordered then
$S_{\rm z}<S_{\rm u}$ requiring $\partial g_c/\partial T <0$.  

\subsection{Specific heat}
\label{sec:specific-heat}
The specific heat relation  for $g=0$ follows from Eq. \eqref{eq:11} as
\begin{equation}
  \label{eq:16}
   C_{\rm z}(T_c,0)-C_{\rm u}(T_c,0)= T\; \chi\left( \frac{\partial
       g_c(T)}{\partial T}\right)_{g=0}^2,
\end{equation}
where $\chi$ is the extensibility of the unzipped chain at the melting point $T=T_c$.
Eq. \eqref{eq:16} gives the discontinuity in the specific heat expected at the melting
point, provided $\partial g_c/\partial T$ is finite.  If the entropy
change is finite, there is a latent heat which contributes a
$\delta$-function peak at the transition point. Eq. \eqref{eq:16} is a
special case of the general formula valid for all $g$, viz.,
\begin{eqnarray}
  \label{eq:42}
 \lefteqn{  C_{\rm z}(T,g_c)-C_{\rm u}(T,g_c)}\nonumber\\
&=& T \left[\chi\left( \frac{\partial
       g_c}{\partial T}\right)^2%\nonumber\\
    + x(g_c)  \frac{\partial^2
       g_c}{\partial T^2} + \frac{\partial \chi}{\partial T} g_c \frac{\partial
       g_c}{\partial T}\right],
\end{eqnarray}
with an extra latent heat contribution.  The derivatives appearing in
Eq. \eqref{eq:42} may conspire to make the RHS zero.  The specific
heat curve will then have only a delta function  at the transition
point superposed on a continuous specific heat curve.

\subsection{Phase boundary}
\label{sec:phase-boundary}

%%%%%%%%%%%%%%%
\figkappa
%%%%%%%%%%%%%%%

The shape of the unzipping phase boundary near
the zero force melting point can be described asymptotically by  (Fig. \ref{fig:3}) 
\begin{equation}
  \label{eq:12}
  g_c(T)\sim |T-T_c|^{\kappa}, \ 
{\rm for}\ g_c(T)\to 0,\ T\to T_c.
\end{equation}
Depending on the value of $\kappa$, a few cases can be considered as
$g_c \partial g_c/\partial T\sim |T-T_c|^{2\kappa-1}$.
\begin{enumerate}
\item If $\kappa>1/2$, then $\frac{\partial g_c(T)}{\partial T}$ remains finite.
 At the zero force melting
  point, there is no change in  entropy  or no latent
  heat.  In this situation, the melting transition is continuous.
\item If $\kappa = 1/2$, there is a latent
  heat and the melting transition is first order.

\item Since infinite latent heat is not possible, there is  a strict
  lower  bound: $\kappa\geq 1/2$.
\end{enumerate}
The shape of the phase boundary, as determined by the exponent
$\kappa$, is linked to the order of the melting transition.

%%%%%%%%%%%%%%%%
\figextr
%%%%%%%%%%%%%%%

Away from melting, in general, the right hand side of Eq.
\eqref{eq:11} is not zero, unless $\partial g_c/\partial T =0$.  The
force induced unzipping transition is necessarily first order.  The
extremum of the phase boundary (Point C at $(T_m, g_m)$ in Fig.
\ref{fig:4}) is a special case.   In absence of any nonanalyticity in
the phase boundary, both phases will have same entropy but with a
discontinuity in the specific heat as per Eq. \eqref{eq:42}.  Since
both $g$ and $T$ are intensive variables, every point in the $T$-$g$
plane represents a unique phase of the DNA, except on the transition
line.  Along a path ACB, there is no real change in phase and no
latent heat is expected.  There is however the possibility of
occurrence of the zipped phase at C.  One may therefore measure either
$C_{\rm u}$ or $C_{\rm z}$.  If $T$ is kept constant at $T_m$,
specific heat will show a discontinuity as we cross C in the phase
diagram vertically.  This looks like a continuous transition.

When $\partial g_c/\partial T >0$, the unzipped phase becomes more
ordered than the zipped phase.   This counter-intuitive behaviour is an
example of a re-entrant phase transition.
It   occurs because the unzipping force acts as stretching forces on
the two unbound chains, orienting them at low temperatures in the
direction of the force reducing the entropy, while the flexibility of
the zipped phase, because of the
impenetrability of the force,  contributes to the 
entropy.

\section{Comparison with Exact results}
\label{sec:exact-results}

There are several models for which exact solutions for the unzipping
transition are known.  We compare the thermodynamics results  with a few such cases.
\subsection{Continuum Gaussian models}
\label{sec:cont-gauss-models}
The unzipping transition was first proved in Ref. \cite{smb1,smb2} for
Gaussian polymers interacting with same monomer index as in DNA.  The
transition line in $(d+1)$-dimensions is given by $g_c(T)\sim
|T-T_c|^{1/(d-2)}$, i.e., $\kappa=\frac{1}{d-2}$ for $2\le d\le 4$.
The zero force melting is continuous but the unzipping transition is
first order.  For $d>4$, the melting transition is first order and 
there is a $\kappa=1/2$ behaviour, as also
found in the lattice modes of Ref. \cite{MTM2001} discussed below.

The model showed that the bound, zipped state does not allow the force
to penetrate and after the unzipping transition the strands are
stretched by the pulling force.  $\partial g_c/\partial T <0$.

A necklace  model analysis shows that $\kappa$ is determined by the
size exponent of the polymer
provided there is no other length scale, i.e.,  $\kappa=\nu$,
where $\nu$ is the size exponent\cite{orland2001,kafri2002}. 
For a first order melting point, since all other length scales remain finite
the relevant length scale is the size of the polymer.  For
Gaussian polymers $\nu=1/2$, giving $\kappa=1/2$, as we saw above for $d>4$.
For continuous melting.  the thermal correlation length is going to
play an important role, giving a different $\kappa$.

\subsection{ Lattice models with bubbles: continuous melting}
\label{sec:with-bubbl-cont}

The unzipping transition problem  can be solved exactly for  
a class of lattice models involving directed polymers in $d+1$
dimensions\cite{marensmb,MTM2001}. 
For the model with bubbles,   there is a continuous melting transition
in dimensions $d<4$ as for the continuum case.  

\subsubsection{d=1 }
\label{sec:d=1-}
For the $1+1$ dimensional model if the two strands are not allowed to
cross, the free energies are\cite{marensmb}
\begin{subequations}
\begin{eqnarray}
  \label{eq:14}
  G_{\rm z}(T,g)&=&k_B T \ln z_{\rm z}(T),\\
  z_{\rm z}(T)&=&\sqrt{1-e^{-\beta}}-1+e^{-\beta},\\ 
  G_{\rm u}(T,g)&=&k_BT\ln z_{\rm u}(T,g), \label{eq:43}\\
 z_{\rm u}(T,g)&=&[2+ 2\cosh (\beta g)]^{-1},
\end{eqnarray}
\end{subequations}
where $\beta=(k_BT)^{-1}$. We choose $k_B=1$. Here $G_{\rm z}$ is independent of 
$g$ because of the impenetration of the force.
The melting transition ($g=0$) at $T_c=[\ln(4/3)]^{-1}$ is continuous with a finite
discontinuity of the specific heat. 

The unzipping phase boundary is given by 
\begin{equation}
  \label{eq:15}
  g_c(T)=T \cosh^{-1}(p(\beta) -1),\quad p(\beta)=(2 z_{\rm z})^{-1},
\end{equation}
obtained by equating the two free energies at the unzipping
transition, i.e., from $G_{\rm z}(T,g_c)=G_{\rm u}(T,g_c)$. Close to
$T_c$ where $z_{\rm u}\to 1/4$, 
and $g_c\to 0$.  The shape  is
\begin{equation}
  \label{eq:23}
  g_c(T)\approx \frac{2  e^{-1/{T_c}}}{\sqrt{1-e^{-1/{T_c}}}{T_c}}\ (T_c-T),
\end{equation}
i.e., $\kappa=1$ (see Fig. \ref{fig:3}).

The extensibility comes from the derivative of $G_{\rm u}$ as
\begin{equation}
  \label{eq:36}
\chi=\frac{1}{2T}\; {\rm sech}^2(g/2T). 
\end{equation}
Linear response is expected in the small force limit, when
$x= g/(2T)$.
At the transition
\begin{equation}
 x_c=\tanh \left(\frac{g_c}{2 T}\right).
\end{equation}
The free energy near an unzipping point $(T,g_c)$  can be written as 
\begin{eqnarray}
  \label{eq:17}
  G_{\rm z}(T,g)&=&  G_{\rm u}(T,g) - T\; \ln \frac{z_{\rm u}(T,g)}{z_{\rm z}(T)}\\
                &=&  G_{\rm u}(T,g) - T\; \ln \frac{z_{\rm u}(T,g)}{z_{\rm u}(T,g_c)},
\end{eqnarray}
by  the continuity of the free energy at the transition point.
Close to the melting point $T=T_c$, $g_c$ 
is small.
In this region, for a small $g$,  an expansion gives
\begin{equation}
  \label{eq:19}
  G_{\rm z}(T,g)=  G_{\rm u}(T,g) +\frac{1}{2} \chi(T_c) (g^2 -g_c^2),
\end{equation}
consistent with Eq. \eqref{eq:20} based on thermodynamic work.

For specific heat,  the discontinuity at $T=T_c$ for
$g=0$ is just the specific heat of the bound state because the unbound
state at $g=0$ has zero specific heat.  A differentiation of
Eq. \eqref{eq:14} shows the agreement with  
the RHS of Eq. \eqref{eq:16}.  Ref. \cite{kapri2006} shows the
behaviour of specific heat for a force that shows reentrance. 
For a nonzero force, the latent heat from Eq. \eqref{eq:11} can
be verified directly.
The phase diagram shows reentrance and an extrema as in
Fig. \ref{fig:4}, recovering the
features discussed in the previous section.
Fig. \ref{fig:6} shows the specific heat as we go through the peak C
in the vertical direction keeping $T=T_m$ and horizontally by keeping
$g=g_m$.   The entropy is continuous. The
specific heat shows a discontinuity along the vertical direction.
Along the horizontal direction of the phase diagram, the entropy is
continuous but the specific heat has one single point for the zipped
phase.  There is no identifiable critical region.
The results are fully consistent with our discussions in Sec. II.  

\figsp

\subsection{Y-model: first order melting}
\label{sec:y-model:-first}

A model of DNA that does not allow any bubble is also exactly
solvable\cite{marensmb,MTM2001}. The thermal melting corresponds to 
an all or none type
behaviour, all base pairs are either formed or broken.  In the bound
state, the number of configurations is $2^N$ for $N$ bonds, while it
is $2^N$ for each strand in the unzipped state.  The free energies are
of the form of Eqs. \eqref{eq:14}, \eqref{eq:15}, except
\begin{eqnarray}
  \label{eq:22}
  p(\beta)=\exp(-\beta).
\end{eqnarray}
The all-or-none melting transition is first order with a latent heat
at $k_BT_c=1/\ln 2$.

Near the melting at $p(\beta)=2$, 
the phase boundary behaves as $g_c(T)\approx 2 \sqrt{T_c-T}$, matching with
$\kappa=1/2$ behaviour of Fig. \ref{fig:3} for a first order transition.
Eq. \eqref{eq:19} is valid with appropriate $T_c$ and $g_c$.
Other relations like specific heat, entropy and latent heat can be
directly verified.

\subsection{Other special cases}
\label{sec:other-special-cases}

Ref. \cite{MTM2001} considers several exactly solvable
models. Reentrance is observed in all situations considered except for
the case of two strands with crossing in $1+1$ dimensions.
In this case $\partial g_c/\partial T>0$ for all $T$ with
$T_c\to\infty$.
Therefore, as in Eq. \eqref{eq:11}, by increasing temperature under
a force, it is possible to get a double stranded bound state at high
temperatures from an unzipped state.

Another special situation is the $2+1$ dimensional model without
crossing of chains.  The phase boundary is $g_c(T)\sim
\exp[-a/(T-T_c)]$ 
with $\partial g_c/\partial T\to 0$ as $T\to T_c$.  
It is possible to unzip
for $T$ close to  but below $T_c$ by an arbitrarily small force.  
Even though this case does not correspond to the power law form of
Eq. \eqref{eq:12}, 
the thermodynamic relations can still be verified including the reentrance
behaviour and the behaviour near the extrema of the phase boundary.

\subsection{Summary of model comparison}
\label{sec:summ-model-comp}
As mentioned  exact results are available for a large class of models in
various dimensions.   In all these cases,  we  find that the general
thermodynamic predictions {\it based on 
impenetrability of the force} below the unzipping transition are
consistent with the extant results.  This lends credence to a general
thermodynamic analysis of various thermodynamic functions near the
transition and phase boundary.

\section{Continuous unzipping transition}
\label{sec:cont-unzipp-trans}
Thermodynamics, by itself, does not exclude the possibility of a continuous
transition under force of the type shown in Fig. \ref{fig:5}(b). In
this section we obtain the thermodynamic relations for such a
continuous transition.  We here assume that the force affects the bound state 
over a range $g_{c1}\le g\le g_{c2}$, but leaves the bound state as it is for 
smaller forces.  In other words, a DNA in its bound state is resilient to small 
forces but allows it to penetrate and alter its nature over a range of forces.

Instead of a jump discontinuity in the isotherm, we allow a continuous
transition at a force $g_{c1}$ where the DNA goes from the bound to a
phase different from the unzipped phase.  The intermediate phase is
further assumed to undergo a transition to the unzipped phase at
$g_{c2}$.  In case, $g_{c2}\to \infty$, there is only one phase in the
high force regime.  However, since single stranded DNA under a force
is a stable thermodynamic system, we expect $g_{c2}$ to remain finite.
There is therefore a range of forces $g_{c1}<g<g_{c2}$ for which the
DNA is affected in a nontrivial way by the force.  Such a scenario was
considered in Ref. \cite{mixed}.  The intermediate state is called a
mixed state.

Near the two transition points, we take the  
isotherm to behave  as 
\begin{equation}
  \label{eq:18}
  x \approx \left\{\begin{array}{lcc}
                    a_1 \;(g-g_{c1})^{\beta},&{\rm for}&  g\to g_{c1}+,\\ 
                  \chi\; g - a \;(g_{c2}-g), &{\rm for}& g \to g_{c2}-,
                    \end{array}\right.
\end{equation}
with $\beta,a,a_1>0$ ($\beta$ is not to be confused with $1/k_BT$).
The unzipped phase for $g>g_{c2}$ is taken, for simplicity, to be in
the linear response regime, $x=\chi g$.  
The exponent $\beta$, by universality,  is same along the transition
line.
 
With the help of the formula for work at a constant temperature from
$g$ to $g_{c2}$, the Gibbs free energy, Eq. \eqref{eq:9}, can be
written as
\begin{eqnarray}
  \label{eq:24}
  G_{\rm m}(T,g)&=& G_{\rm m}(T,g_{c2})+ \frac{1}{2} \chi
  (g_{c2}^2-g^2)\nonumber\\
&&\quad - \frac{1}{2} a\;(g_{c2}-g)^2,
\end{eqnarray}
where the subscript m indicates the mixed or the intermediate state.
By continuity, $G_{\rm m}(T,g_{c2})=G_{\rm u}(T,g_{c2})$, and
therefore,
\begin{equation}
  \label{eq:25}
  G_{\rm m}(g,T)= G_{\rm u}(g,T) - \frac{1}{2} a\;(g_{c2}-g)^2,
\end{equation}
for $g$ close to but smaller than $g_{c2}$.
This form not only shows that the mixed state has a lower free energy than the
unzipped state, but also gives the specific heat behaviour at
$g_{c2}$, as (see Eq. \eqref{eq:33})
\begin{equation}
  \label{eq:26}
  C_{\rm m}(T,g_{c2})-  C_{\rm u}(T,g_{c2}) = -T a \left(\frac{\partial
    g_{c2}}{\partial T}\right)^2.
\end{equation}

The specific heat relation derived in the appendix is also applicable
for the z to m transition.   In this case the zipped phase has zero 
extensibility and therefore
\begin{equation}
  \label{eq:35}
    C_{\rm m}(T,g_{c1})-  C_{\rm z}(T,g_{c1}) = - T  \left(\frac{\partial
    g_{c1}}{\partial T}\right)^2 a_1 \beta (g-g_{c1})^{\beta-1},
\end{equation}
indicating the possibility of a diverging specific heat if $\beta<1$.

\section{Conclusion}
\label{sec:conclusion}
In this paper the thermodynamic description of the DNA unzipping phase
transition is discussed.  Without considering any microscopic details,
we show that the thermodynamic relations in the fixed force ensemble
have all the important features of the phase transition.  Here we
concentrate only on the force induced unzipping by pulling the two
strands apart.  A linear response has been used for evaluating the
work by force, but the analysis can be carried out keeping the full
form.  Although a first order phase transition is observed in various
models, the possibility of a continuous transition did not get much
attention. Thermodynamics does not discard this possibility, and hence
we extend our study to the case of continuous transition. The only
information we use as an input to our analysis is that the zipped
phase does not allow the force to penetrate below a certain critical
force in the first order phase transition. The behaviour of the change
in entropy and the specific heat go in accordance with the observed
features in some known models. Various cases of the phase boundary
line near the melting point are also analyzed. For the continuous
transition there is an additional region in the phase diagram, showing
a possible mixed phase, which allows the force to penetrate.  We
proceed with the general forms of the isotherms.  This phenomenon of
partial penetration of force looks very much like type II
superconductors.  Even though the variables and the microscopic
origins are different in the two cases, there is a striking similarity
between the relations obtained here for DNA and thermodynamic
relations for superconductors. Lastly, we restricted ourselves to the
unzipping force here but similar analysis can be done for the other
forces like stretching force and pressure.  It would be interesting to
observe the effect of these forces on such transitions.

\appendix

\section{Other external forces}
\label{sec:other-extern-forc}

As discussed in the Introduction, there could be other forces like
pressure and the stretching force.
To include  these  external forces, the Gibbs free energy of
Eq. \eqref{eq:1} needs two additional terms 
\begin{equation}
\label{eq:37}
  G(T,g,P,f)=U-T\;S\;-\;g\;x\;+\;P\;V\;-\;{\bf f\cdot R},
\end{equation}
where $V$ refers to the volume and ${\bf R}={\bf R}_1+{\bf R}_2$ to
the end to end distances of the chains.  In this notation \mbox{$x=|{\bf
  R}_1-{\bf R}_2|$} 
with the subscript denoting the two chains.  The conjugate variable is
volume ($V$), but it is the volume of the polymer with the surrounding
distorted solvent layers.
For simplicity we ignored the terms
involving ${\bf f}$ and $P$, and considered only the unzipping force
$g$.  The extended form of the free energy shows that it is possible
to have cross-effects like the $f$ and $P$ dependence of $x$.  
  
Different studies looked at the effect of hydrostatic pressure, e.g.,
the stability of hairpins, B-DNA, the stalling of transcription
elongation
complexes\cite{amiri,wiltonBDNA,chalikian,takahashi2013,erijman,robinson94}.
The melting temperature $T_c$ seems to depend on the hydrostatic
pressure, $P$, only at a very high $P$.  It is possible to measure the
adiabatic compressibility (constant entropy) of a DNA molecule by
measuring the velocity of ultrasonic waves.  A different way of
exerting a pressure is to use osmolytes like polyethylene glycol (PEG)
and other molecules that cannot penetrate the DNA.  There are reports
of hydrostatic pressure reversing the effect of osmotic pressure in
protein-DNA interaction.

There can be a stretching force ($f$) that distorts the shape and
tries to elongate the chain.  What one finds is a transformation of a
dsDNA to an ``overstretched'' state with its length increasing by a
factor of 1.7 \cite{smith1996,overstretch,senooverstret}.  Whether it
is an thermodynamic (meaning equilibrium) transition is still debated.
The conjugate variable is the end-to-end distance (${\bf R}$), if the
end points are tied together.  The conjugate variable becomes the
length of the polymer for the overstretching transition.  This is the
usual force considered for a polymer\cite{degennes,smbflory}, but
coupling to unzipping seems to lead to the new feature of
overstretching.  As a perturbation to an entropy-dominated polymer
configuration, the response to the stretching force need not be
linear, if the chain does not behave as Gaussian, but the
overstretching transition is beyond this regime where the finite
extensions of the bonds need to be taken into account.  There are
evidences of overstretching being coupled with unzipping making cross
terms important.  The response functions needed for such cross effects
would be $\chi_i=\partial x/\partial f_i, \chi_P=\partial x/\partial
P$ with other appropriate variables kept constant.

\section{Maxwell relations}
\label{sec:maxwell-relations}

The differential relations of the free energy are of the expected type
\begin{eqnarray}
  \label{eq:2}
  dU&=&\ \ T\; dS + g\;dx,\\
  dF&=&-S\;dT+g\;dx,  \label{eq:38}
\end{eqnarray}
In the canonical fixed-$g$ case, the conjugate parameters and the
response functions are the
first and second derivatives respectively of the appropriate
thermodynamic potential as
\begin{subequations}
\begin{eqnarray}
  S=-\left.\frac{\partial G}{\partial T}\right|_{g},& &
 \frac{1}{T} C_g=\left.\frac{\partial S}{\partial
     T}\right|_{g}=-\left.\frac{\partial^2 G}{\partial
     T^2}\right|_{g}, \label{eq:4}\\
  x=-\left.\frac{\partial G}{\partial g}\right|_{T} 
& &\chi_T=\left.\frac{\partial x}{\partial
     g}\right|_{T}=-\left.\frac{\partial^2 G}{\partial
      g^2}\right|_{T},\label{eq:5}
\end{eqnarray}
where $C_g$ is the constant force heat capacity and $\chi_T$ is the
extensibility at constant temperature, the local slope of a $g$-$x$ isotherm. 
As usual, the positivity of $C_g$ and $\chi_T$, needed for
stability, are related to the convexity conditions satisfied by $G$.

The differential forms  yield the  Maxwell relations for DNA as
\begin{eqnarray}
  \label{eq:6}
  \left.\frac{\partial x}{\partial T}\right|_{g}=  \left.\frac{\partial
      S}{\partial g}\right|_{T}, &\quad {\rm and}&   \left.\frac{\partial x}{\partial T}\right|_{S}=  \left.\frac{\partial S}{\partial g}\right|_{x},\\
  \left.\frac{\partial S}{\partial x}\right|_{T}=  -\left.\frac{\partial
      g}{\partial T}\right|_{x}, &\quad {\rm and}&   -\left.\frac{\partial T}{\partial g}\right|_{S}=  \left.\frac{\partial x}{\partial S}\right|_{g},\label{eq:32}
\end{eqnarray}
\end{subequations}
of which the first two relate the thermal expansion of the open fork
to the heat flow for change in force.

\section{Specific heat near a line of continuous transition}
\label{sec:specific-heat-near}

Let us consider a phase boundary $g=g^*(T)$, where at any point
$(T,g^*(T))$, the Gibbs free energies and the entropies of the two
phases A and B are the same, i.e.,
\begin{eqnarray}
  \label{eq:27}
  G_{\rm A}(g^*,T)&=&  G_{\rm B}(g^*,T),\\
 \vspace{-4cm}{\rm and}&&\nonumber\\ 
  S_{\rm A}(g^*,T)&=&  S_{\rm B}(g^*,T).
\end{eqnarray}
Along the phase boundary, at a neighbouring point, $G_{\rm
  A}(g^*+dg^*,T+dT)=G_{\rm B}(g^*+dg^*,T+dT)$.  An expansion gives
\begin{equation}
  \label{eq:28}
  \frac{\partial G_{\rm A}}{\partial T} \;dT +  \frac{\partial G_{\rm
      A}}{\partial g^*} \;dg^*
= \frac{\partial G_{\rm B}}{\partial T} \;dT +  \frac{\partial G_{\rm
      A}}{\partial g^*} \;dg^*,
\end{equation}
which tells us that at the transition point, the conjugate variable
$x$ is continuous, because of the continuity of the entropy
($S=-\partial G/\partial T$).

The constant force specific heat is given by 
\begin{equation}
  \label{eq:29}
  C_{\rm A}=T \left.\frac{\partial S_{\rm A}}{\partial T}\right |_g.
\end{equation}
The derivative of the entropy can be expressed in terms of the
derivative along the transition line as
\begin{equation}
  \label{eq:30}
  \frac{d S_{\rm A}}{d T\phantom{x}}= \left.\frac{\partial S_{\rm
        A}}{\partial T}\right |_g + \left.\frac{\partial S_{\rm
        A}}{\partial g}\right|_T\; \frac{\partial g^*}{\partial T},
\end{equation}
and a similar relation for phase B.  Since at each point on the
transition line entropy is continuous, $\frac{d S_{\rm A}}{d
  T\phantom{x}}=\frac{d S_{\rm B}}{d T\phantom{x}}$.
Eq. \eqref{eq:29} can now be used to express the constant force specific heat
difference as
\begin{equation}
  \label{eq:31}
  C_{\rm A}-C_{\rm B}= T  \frac{\partial g^*}{\partial T}
  \left[\left.\frac{\partial S_{\rm  A}}{\partial g}\right|_T\;
  -\left.\frac{\partial S_{\rm B}}{\partial g}\right|_T\;\right].
\end{equation}
A further simplification can be achieved by using one of the Maxwell
relations, Eq. \eqref{eq:32}, 
\begin{equation}
  \label{eq:34}
  \left.\frac{\partial S_{\rm A}}{\partial g}\right|_T = 
  \left.\frac{\partial S_{\rm A}}{\partial x}\right|_T \;
  \left.\frac{\partial x}{\partial g}\right|_T=  \frac{\partial
    g^*}{\partial T} \;  \chi_A, 
\end{equation}
where $\chi_{\rm A}$ is the extensibility of phase A.  With a similar
relation for phase B, we obtain
\begin{equation}
  \label{eq:33}
  C_{\rm A}-C_{\rm B}= - T  \left(\frac{\partial g^*}{\partial
      T}\right)^2 \left(\chi_{\rm A}-\chi_{\rm B}\right ).
\end{equation}
If the extensibility of phase A is less than that of B, then the 
specific heat of phase A is higher than that of B.

\end{document}